\documentclass[twocolumn]{aa}

\usepackage{amsmath}
\usepackage[varg]{txfonts}
\usepackage{graphicx}
\usepackage{epstopdf}
\usepackage[draft]{hyperref}
\usepackage{ulem}

\institute{Hamburger Sternwarte, Universit\"at Hamburg, Gojenbergsweg
  112, 21029 Hamburg, Germany \label{inst1}
  \and
  Departamento de Astronom\'ia, Facultad Ciencias F\`isicas y Matem\`aticas, Universidad de Concepci\'on, Av. Esteban Iturra s/n Barrio Universitario, Casilla 160-C, Concepci\'on, Chile \label{inst2}}

\title{Eclipsing time variations in close binary systems: Planetary hypothesis vs. Applegate mechanism}
\titlerunning{Planetary hypothesis vs. Applegate mechanism}
\author{M. V\"olschow \inst{\ref{inst1}, \ref{inst2}}
\and D. R. G. Schleicher \inst{\ref{inst2}}
\and V. Perdelwitz \inst{\ref{inst1}}
\and R. Banerjee \inst{\ref{inst1}}}
\authorrunning{V\"olschow et al.}

\offprints{{\tt marcel.voelschow@hs.uni-hamburg.de}}
\date{}

\def\D{\mathrm{d}}

\newcommand{\s}{\rm{s}}

\newcommand{\erg}{\rm{ erg}}
\newcommand{\Msol}{{\rm ~M}_{\sun}}
\newcommand{\Rsol}{{\rm ~R}_{\sun}}
\newcommand{\Lsol}{{\rm ~L}_{\sun}}
\newcommand{\Mjup}{{\rm ~M}_{Jup}}
\newcommand{\pc}{\rm{ pc}}

\newcommand{\au}{\rm{au}}
\newcommand{\K}{\rm{ K}}

\newcommand{\yr}{\rm{yr}}
\newcommand{\days}{\rm{d}}

\newcommand{\Myr}{\mbox{ Myr}}
\newcommand{\Gyr}{\mbox{ Gyr}}

\newcommand{\fsequa}{~~ .}
\newcommand{\comequa}{~~ ,}

\newcommand{\bef}{\begin{figure}[!t]}
\newcommand{\eef}{\end{figure}}

\abstract{
The observed eclipsing time variations in post-common-envelope binaries (PCEBs) can be interpreted as potential evidence for massive Jupiter-like planets, or as a result of magnetic activity, leading to quasi-periodic changes in the quadrupole moment of the secondary star. The latter is commonly referred to as the Applegate mechanism. Following \citet{Brinkworth2006}, we employ here an improved version of Applegate's model including the angular momentum exchange between a finite shell and the core of the star. The framework is employed to derive the general conditions under which the Applegate mechanism can work, and is subsequently applied to a sample of $16$ close binary systems with potential planets, including $11$ PCEBs. Further, we present a detailed derivation and study of analytical models which allow for an straightforward extension to other systems. Using our full numerical framework, we show that the Applegate mechanism can clearly explain the observed eclipsing time variations in 4 of the systems, while the required energy to produce the quadrupole moment variations is too high in at least 8 systems. In the remaining 4 systems, the required energy is comparable to the available energy produced by the star, which we consider as borderline cases. Therefore, the Applegate mechanism cannot uniquely explain the observed period time variations for this entire population. Even in systems where the required energy is too high, the Applegate mechanism may provide an additional scatter, which needs to be considered in the derivation and analysis of planetary models.
}  


\begin{document}

\maketitle

\section{Introduction}
\label{sec:intro}

Short period binaries with dwarf components have proven to be ideal systems to constrain stellar structure and evolution models with precise mass and radius data. White Dwarf / Red Dwarf (hereafter WD+RD) binaries are of particular relevance, as their large difference both in radius and luminosity results in accurate lightcurves very close to ideal predictions, allowing for high precision timing measurements \citep[see, e.g.,][]{Parsons2010}.

Such binary systems form as a result of a common envelope (CE) phase. The CE model was originally put forward by \citet{Paczynski76}, and assumes that the primary star in an originally wider ($\sim1$~AU) binary system  has reached the giant branch, with the secondary being engulfed by the envelope of the giant. As a result of this, both the energy and angular momentum of the secondary are deposited into the envelope of the giant, causing the secondary to spiral inward towards scales of about a solar radius. These models have been refined in subsequent studies, e.g. \citet{Meyer79, Iben93, Taam00, Webbink08} and \citet{Taam10}. The CE phase has been simulated both using smoothed-particle-hydrodynamics (SPH) and grid-based techniques by \citet{Passy12}, and a review of the current state of the art is provided by \citet{Ivanova13}. The systems resulting from such a CE phase consist  of a White Dwarf resulting from the core of the giant and a Red Dwarf on a narrow orbit of $\sim1$~$R_\odot$. They are referred to as post-common-envelope binaries (PCEBs).

The evolution of close binary systems is predominantly governed by angular momentum loss, which can be driven by gravitational wave emission for binaries with very short periods ($P_{\rm orb}<3$~h) \citep{Kraft1962, Faulkner1971} or magnetic breaking for binaries with $P_{\rm orb}>3$~h \citep{Verbunt1981}. While the resulting angular momentum loss implies a continuous decrease of the orbital period over time, a number of systems is known showing regular quasi-periodic modulations of the order $\Delta P/P_{bin}\sim3\times10^{-5}$ with periods of several decades in Algols, RS Canum Venaticorum (RS CVn), W Ursae Majoris and cataclysmic variable (CV) stars \citep{Ibanoglu91}. These variations are studied employing O-C diagrams, in which O denotes the observed orbital phase of the binary at a given time, from which a correction C is subtracted based on the zero- and first-order terms in the expansion of the angular velocity, i.e. assuming $\omega(t)=\omega_0+\dot{\omega}t+..$. A diagram showing O-C vs time provides information on the quasi-periodic modulations, corresponding to fluctuations including a regular increase accompanied by a subsequent decrease in the orbital period. The period variation is related to the amplitude in the O-C diagram via \citep{Applegate1992}

\begin{equation}
\frac{\Delta P}{P_{bin}}=2\pi \frac{O-C}{P_{\rm mod}},
\end{equation}

where $P_{\rm mod}$ denotes the period of the modulation. In the case of eclipsing binaries, these fluctuations can be accurately measured using transit timing variations (TTVs) to constrain the underlying physical mechanism.

A potential mechanism to explain the eclipsing time variations has been proposed by \citet{Applegate1992}, in which the period variations are explained as a result of quasi-periodic changes in the quadrupole moment of the secondary star as a result of magnetic activity. It is thus assumed that a sufficiently strong magnetic field is regularly produced during a dynamo cycle, leading to a redistribution of the angular momentum within the star and therefore a change in its quadrupole moment. This model was motivated from a sample of $101$ Algols studied by \citet{Hall1989}, showing a strong connection between the orbital period variations and the presence of magnetic activity.

To sufficiently change the stellar structure to drive the quasi-periodic period oscillation, a certain amount of energy is required to build up a strong magnetic field, which is subsequently dissipated and again built up in the next dynamo cycle. Ultimately, such energy should be extracted from the convective energy of the star, which is powered by the nuclear energy production \citep{Marsh90}. While sufficient energy appears to be available in the case of Algols, it has been shown by \citet{Lanza2005} that the Applegate scenario needs to be rejected for RS~CVns, and independently \citet{Brinkworth2006} have reported that the orbital period variations of the PCEB system NN~Ser cannot be explained using Applegate's model.

In fact, the orbital period variations in NN~Ser were shown to be consistent with a two-planet solution, which was shown to be dynamically stable \citep{Beuermann2010, Horner2012, Beuermann2013, Marsh14}. Similarly, the planetary solution of HW~Vir was shown to be secularly stable \citep{Beuermann2012b}, while a final conclusion on HU~Aqr \citep{Qian2011, Gozdziewski2012, Hinse2012, Wittenmyer2012, Gozdziewski2015HUAqr} is pending. The eclipsing time variations in QS~Vir, on the other hand, are currently not understood, and appear to be incompatible both with the planetary hypothesis as well as the Applegate scenario \citep{Parsons2010}. A summary of PCEB systems potentially hosting planets has been compiled by \citet{Zorotovic2013}, including the properties of the planetary systems. The  latter correspond to massive giant planets of several Jupiter masses, on planetary orbits of a few AU.


However, the discussion whether the planets are real is still ongoing. In this respect, the proposed brown dwarf in V471~Tau has not been found via direct imaging \citep[][but see comments by \citet{Vaccaro2015}]{Hardy2015}. In this paper, we in fact show that the period time variations in V471~Tau could also be produced by an Applegate mechanism. 

The original Applegate model linking magnetic activity to period time variations has subsequently been improved by different authors. For instance, the analysis by \citet{Lanza1998} provided an improved treatment of the mechanical equilibrium, including the impact of rotational and magnetic energy on the quadrupole moment, leading to an improved estimate of the energy requirements. While the original framework was based on a thin-shell approximation, the latter was extended by \citet{Brinkworth2006} considering a finite shell around the inner core, as well as the change of the quadrupole moment of the core due to the exchange of angular momentum. An even more detailed formulation has been provided by \citet{Lanza2006}. It was  proposed by \citet{Lanza1999} that an energetically more favorable scenario may occur in the presence of an $\alpha^2$ dynamo, which was however shown to require strong assumptions concerning the operation of the dynamo, including a dynamo restricted to the star's equatorial plane and a magnetic field of $10^5$~G \citep{Ruediger2002}.

In this paper, we  adopt the formulation by \citet{Brinkworth2006} and apply it to realistic stellar profiles. Based on this analysis, we  systematically assess whether the Applegate mechanism is feasible in a sample of 16 close binary systems with potential planets, including 11 PCEBs. We  further provide analytical scaling relations which can reproduce our main results, provide further insights into the Applegate mechanism and allow for an extension of this analysis to additional systems. 

\section{Systems investigated in this paper}
\label{sec:systems}

\subsection{Classification}

Our sample consists of a total of $16$ close binary systems with observed period variations potentially indicating the presence of planets. $11$ of these are the PCEB systems previously described by \citet{Zorotovic2011} and \citet{Zorotovic2013}, as well as four RS~CVn binaries RU~Cnc, AW~Her \citep{Tian2009}, HR~1099 \citep{Fekel1983, Garcia2003, Frasca2004} and SZ~Psc \citep{Jakate1976, Popper1988, Wang2010}, and the RR-Lyr type binary BX~DRA \citep{Park2013}. The detailed properties of these systems, including binary type and spectral class, are given in Table~\ref{tab:secondarys}.

\begin{table*}
\caption{Summary of relevant system parameters for the close binaries with planetary candidates investigated in this paper where we use the usual nomenclature and $\tau$ 
beeing the age of the binary system (see also sec.~\ref{sec:systems}). Sources for the data are given below.}
\begin{center}
\begin{tabular}{c|c|c|c|c|c|c|c|c|c|c}
\hline
System & $a_{\rm bin}$/$\Rsol$ & $P_{\rm bin}$/$\days$ &$M_{\rm sec}$/$\Msol$ & $R_{\rm sec}$/$\Rsol$ & $T_{\rm sec}$/$\K$ & $L_{\rm sec}$/$\Lsol$ & $\tau$/$\Gyr$ & type & spectral class & Sources \\
\hline
HS 0705+6700 	& 0.81	& 0.0958 & 0.134 & 0.186 & 2,900 & 0.00219$^{\ast}$ & $5^{**}$ & Al & sdB+dM & 1,2,3 \\
HW Vir 			& 0.860	& 0.117 & 0.142 & 0.175	& 3,084 & 0.003 & $5^{**}$ & Al & sdB+dM & 4,5,6 \\
NN Ser	 		& 0.934	& 0.130 & 0.111 & 0.149 & 2,920 & 0.00147$^{\ast}$ & 2 & No & DA01+M4 & 7,8,9 \\
NSVS14256825	& 0.80	& 0.110 & 0.109 & 0.162 & 2,550 & 0.000994$^{\ast}$ & $5^{**}$ & EB & sd0B+dM & 3 \\
NY Vir		 	& 0.77 & 0.101 & 0.15 & 0.14 & 3,000 & 0.00142$^{\ast}$ & $5^{**}$ & Al & sdB+dM & 3,10,11 \\
HU Aqr		 	& 0.69 & 0.0868 & 0.18 & 0.22 & 3,400 & 0.00580$^{\ast}$ & 1$^{\ast}$ & AM & WD+M4V & 12,13,14 \\
QS Vir		 	& 1.27$^{\ast}$ & 0.151 & 0.43 & 0.42 & 3,100 & 0.0146$^{\ast}$ & $5^{**}$ & Al & DA+M2-4 & 15,16,17 \\
RR Cae		 	& 1.62$^{\ast}$ & 0.304 & 0.183 & 0.209 & 3,100 & 0.00362$^{\ast}$ & 4.5$^{\ast}$ & Al & DA7.8+M4 & 18,19,20 \\
UZ For		 	& 0.788$^{\ast}$ & 0.0879 & 0.14 & 0.177 & 2,950 & 0.00213$^{\ast}$ & 1$^{\ast}$ & AM & M4.5 & 21,22 \\
DP Leo		 	& 0.59$^{\ast}$ & 0.0624 & 0.1 & 0.134$^{\ast}$ & 2,710$^{\ast}$ & 0.000867$^{\ast}$ & 2.5$^{\ast}$ & AM & WD & 23,24 \\
V471 Tau	 	& 3.3 & 0.522$^{\ast}$ & 0.93 	& 0.96 	& 5,040 & 0.40 & 1	& Al & K2V+DA & 25,26,27 \\
\hline
RU Cnc		 	& 27.76$^{\ast}$ & 10.17 & 1.42 & 4.83 & 4,940$^{\ast}$ & 12.5$^{\ast}$ & 3.3$^{\ast}$ & RS & dF9+dG9 & 23,28 \\
AW Her		 	& 24.86 & 8.82$^{\ast}$ & 1.35 & 3.0 & 5,110$^{\ast}$ & 5.49$^{\ast}$ & 4.0$^{\ast}$ & RS & G2IV &23, 28 \\
HR 1099		 	& 11.2$^{\ast}$ & 2.84 & 1.3 & 4.0 & 4,940$^{\ast}$ & 8.6$^{\ast}$ & 4.5$^{\ast}$ &RS & K2	& 29,30,31,32 \\
BX Dra		 	& 4.06 & 0.579 & 2.08 & 2.13 & 6,980 & 9.66 & 0.5$^{\ast}$ & RR & F0IV-V & 23,33 \\
SZ Psc		 	& 15.04$^{\ast}$ & 3.97 & 1.62 & 5.1 & 5,004$^{\ast}$ & 14.7$^{\ast}$ & 2.1$^{\ast}$ & RS & K1IV+F8V & 23,34,35,36 \\
\hline
\end{tabular}
\tablefoot{We marked calculated values with an asteriks. In case no age estimates are given in the literature, we adopt a canonical age of $5~\Gyr$, marked as $^{**}$. The term secondary does not necessarily refer to the lower mass component. Rather, it refers to the component of the binary for which we pursued the calculations. For RR Cae, DP Leo and UZ For, we estimated the WD progenitor masses using the fits provided by \citet{Meng2008} assuming solar metallicity. In the case of DP Leo and UZ For, we adopted the main-sequence lifetime of the progenitor plus $0.5~\Myr$ as a rough age estimate. In RR Cae, we added the cooling age of the WD which is given as $t_{cool} \sim 1 \Gyr$ by ref. 18. The abbreviations for the system type are: Al: eclipsing binary of Algol type (detached), EB: eclipsing binary, No: Nova, AM: CV of AM Her type (polar), RR: variable star of RR Lyr type, RS: variable star of RS CVn type. The horizontal line separates the PCEB systems from other close binaries.}
\tablebib{
(1)~\citet{Beuermann2012a}; (2)~\citet{Drechsel2001}; (3)~\citet{Almeida2012};
(4)~\citet{Beuermann2012b}; (5)~\citet{Lee2009}; (6)~\citet{Wood99};
(7)~\citet{Parsons2010NNSer}; (8)~\citet{Beuermann2010}; (9)~\citet{Beuermann2013};
(3)~\citet{Almeida2012}; 
(10)~\citet{Kilkenny1998}; (11)~\citet{Qian2012};
(12)~\citet{Schwope2011}; (13)~\citet{Wittenmyer2012}; (14)~\citet{Gozdziewski2015HUAqr};
(15)~\citet{ODonoghue2003}; (16)~\citet{Parsons2010}; (17)~\citet{Drake14};
(18)~\citet{Maxted2007}; (19)~\citet{Gianninas11}; (20)~\citet{Zorotovic2013};
(21)~\citet{Bailey1991}; (22)~\citet{Potter2011};
(23)~\citet{Pourbaix04}; (24)~\citet{Beuermann2011};
(25)~\citet{OBrien2001}; (26)~\citet{Hussain06}; (27)~\citet{Hardy2015};
(28)~\citet{Tian2009}; 
(29)~\citet{Fekel1983}; (30)~\citet{Garcia2003}; (31)~\citet{Frasca2004}; (32)~\citet{Gray06};
(33)~\citet{Park2013};
(34)~\citet{Jakate1976}; (35)~\citet{Popper1988}; (36)~\citet{Wang2010}
.}
\label{tab:secondarys}
\end{center}
\end{table*}

\subsection{Period variations and the LTV effect}

In the presence of a planet, the close binary system and the companion revolve around their common barycentre resulting in variations of the observed eclipse timings referred to as the light travel time variation (LTV) effect.

In the binary eclipse O-C diagram, the expected signature of a single planet on a circular orbit is sinusoidal with semi-amplitude $K$ and period $P_{\rm plan}$. Treating the binary as a point mass and further assuming an edge-on binary (i.e., $i=90\deg$), we can approximate the semi-amplitude of the LTV $K$ caused by the accompanying object via

\begin{equation}
K = \frac{ M_{\rm plan} G^{1/3}}{c} \left[ \frac{P_{\rm plan}}{2 \pi (M_{\rm pri}+M_{\rm sec})} \right]^{2/3},
\end{equation}

where $M_{\rm pri}$ and $M_{\rm sec}$ denote the eclipsing binary component masses, $M_{\rm plan}$ the planetary mass, $P_{\rm plan}$ the planetary orbital period, c the speed of light and $G$ the gravitational constant (see \citet{Pribulla2012}  for a more detailed description). Relative to the binary's period, we have a period change of \citep[see, e.g.,][]{Gozdziewski2015}

\begin{equation}
\frac{\Delta P}{P_{\rm bin}} = 4 \pi \frac{K}{P_{\rm plan}} \fsequa
\end{equation}  

In Table~\ref{tab:planets} we summarize the LTV properties of the proposed planetary systems as given in the literature we refer to in Table~\ref{tab:secondarys}.  

\begin{table*}
\caption{Summary of the TTV data. If more than one planet is thought to be present, we only included the planet with the biggest influence on the binary's period, i.e. the largest $\Delta P / P_{bin}$.}
\begin{center}
\begin{tabular}{c|c|c|c}
\hline
System 			& Semi-amplitude $K$/s	& Period $P$/$\yr$ 	& $\Delta P$/$P_{bin}$ \\
\hline
HS 0705+6700	& 	67 		&	8.41	& $3.2\cdot 10^{-6}$ 	\\
HW Vir			& 	563 	&	55		& $4.1\cdot 10^{-6}$ 	\\
NN Ser			& 	27.65 	&	15.482	& $7.1\cdot 10^{-7}$ 	\\
NSVS14256825	& 	20 		&	6.86	& $1.2\cdot 10^{-6}$  	\\
NY Vir			& 	12.2 	&	7.9		& $6.1\cdot 10^{-7}$	\\
HU Aqr			& 	87.7 	&	19.4	& $1.8\cdot 10^{-6}$	\\
QS Vir			& 	43	 	&	16.99	& $10^{-6}$	\\
RR Cae			& 	7.2 	&	11.9	& $2.4\cdot 10^{-7}$	\\
UZ For			& 	21.6 	&	16		& $5.4\cdot 10^{-7}$	\\
DP Leo			& 	16.9 	&	28		& $2.4\cdot 10^{-7}$	\\
V471 Tau		& 	137.2 	&	30.5	& $1.8\cdot 10^{-6}$	\\
\hline
RU Cnc			& 	21.4 	&	37.6	& $2.3\cdot 10^{-7}$	\\
AW Her			& 	1410 	&	12.8	& $4.4\cdot 10^{-5}$	\\
HR 1099			& 	7900 	&	35		& $9.0\cdot 10^{-5}$	\\
BX Dra			& 	532 	&	30.2	& $3.5\cdot 10^{-6}$	\\
SZ Psc			& 	480 	&	55.5	& $3.4\cdot 10^{-6}$	\\
\hline
\end{tabular}
\label{tab:planets}
\end{center}
\end{table*}





\section{Applegate's mechanism}
\label{sec:applegate}

In the following, we introduce the framework to assess the feasibility of Applegate's mechanism. For this purpose, we first remind the reader of the original Applegate framework and the resulting energy requirement to drive quadrupole moment variations of the secondary star. This is contrasted with the finite shell model by \citet{Brinkworth2006}, which considers the angular momentum exchange between the core and a finite shell and accounts for the backreaction of the core. Using this framework, we derive three different approximations of increasing accuracy: An analytical Applegate model assuming a constant density throughout the star, an analytical two-zone model employing different densities in the core and the shell, and a numerical model employing a full stellar density profile. Finally, we present an analytical two-zone model that precisely reproduces the results of the numerical model and leads to new insights into the limitations of the Applegate mechanism.

\subsection{Thin shell model by Applegate (1992)}

We consider a close PCEB system with a binary separation $a_{\rm bin}$ consisting of a White Dwarf, and a Red Dwarf secondary with mass $M_{\rm sec}$ and radius $R_{\rm sec}$. According to \citet{Applegate1992}, the relative period change $\Delta P / P_{bin}$ is related to the secondary's change in quadrupole moment $\Delta Q$ by

\begin{equation}
\frac{\Delta P}{P_{\rm bin}} = -\frac{9 \Delta Q}{a_{\rm bin}^{2} M_{\rm sec}} \fsequa
\end{equation} 

As a result of the secondary's magnetic activity, angular momentum can be transferred between the core and its outer regions leading to a change in their respective angular velocities, causing a change in the oblateness of both the core and the outer shell and modulating the secondary's quadrupole moment. For the sake of simplicity, Applegate considered a thin homogeneous shell and neglected the change of the core's oblateness.\\

We divide the star in an inner core with radius $R_{\rm core}$ and initial orbital velocity $\Omega _1$, and an adjacent outer shell with outer radius $R_{\rm sec}$ and orbital velocity $\Omega _2$, both rotating as rigid objects. Under these assumptions, the required energy $\Delta E$ to drive a given period change via an exchange of angular momentum $\Delta J$ between core and shell is given as

\begin{equation}
\label{eq:applegate_energy}
\Delta E = ( \Omega _2 - \Omega _1 ) \Delta J + \frac{\Delta J ^{2}}{2 I_{\rm eff}},
\end{equation}  

with effective moment of inertia $I _{\rm eff} = I_{\rm core} I_{\rm shell} / (I_{\rm core}+I_{\rm shell}$ \citep[see, e.g.,][]{Applegate1992,Parsons2010}. Assuming a thin shell, one has $I_{\rm eff} = (1/3) M_{\rm shell} R_{\rm core}^{2}$. A given period change $\Delta P$ and an angular momentum exchange are connected via

\begin{equation}
\Delta J = \frac{G M_{\rm sec}^{2}}{R_{\rm sec}} \left( \frac{a_{\rm bin}}{R_{\rm sec}} \right)^{2} \frac{\Delta P}{6 \pi} \fsequa
\end{equation} 

Following \citet{Parsons2010}, one fixes the shell mass and evaluates Eq.~\ref{eq:applegate_energy} for varying core radii and hence angular momentum exchange to solve for the energy minimum.

Combining the framework of \citet{Applegate1992} with an improved model for the variation of the quadrupole moment considering both the rotation and magnetic energy \citep{Lanza1998}, \citet{Tian2009} derived an approximate formula for the relation between the required energy to drive Applegate's mechanism and the observed eclipsing time variations:

\begin{equation}
\label{eq:tian}
\resizebox{\hsize}{!}
{
$\frac{\Delta E}{E_{\rm sec}} = 0.233 \left( \frac{M_{\rm sec}}{\Msol} \right)^{3} \left( \frac{R_{\rm sec}}{\Rsol} \right)^{-10} \left( \frac{T_{\rm sec}}{6000~\K} \right)^{-4} \left( \frac{a_{\rm bin}}{\Rsol} \right)^{4} \left( \frac{\Delta P}{\s} \right)^{2} \left( \frac{P_{\rm mod}}{\yr} \right)^{-1} .$
}
\end{equation}

Both this approximative formula and Applegate's original work only consider angular momentum exchange between a thin outer shell and a dominating core. Moreover, they do not take into account the core's rotational counter-reaction as a result of angular momentum conservation, i.e. an opposite change in the oblateness compensating a fraction of the change of the quadrupole moment. Thus, the energy required to power a certain level of period variation increases and a more realistic description of Applegate's mechanism must include both components. 

\subsection{Finite shell model by Brinkworth et al. (2006)}\label{sec:brinkworth}

In contrast to the original work by \citet{Applegate1992}, \citet{Brinkworth2006} derived an analytic expression for the effective change of the secondary's quadrupole moment $\Delta Q$ for a given change of core rotation $\Delta \Omega _1$ and shell rotation $\Delta \Omega _2$, which reads

\begin{equation}
\label{eq:omega2equa}
\Delta Q =  Q' _1 \, \left[ 2 \Omega _1 \Delta \Omega _1 + ( \Delta \Omega _1 )^{2} \right] + Q' _2 \, \left[2 \Omega _2 \Delta \Omega _2 + ( \Delta \Omega _2 )^{2} \right] \comequa
\end{equation}

where the coefficients $Q' _1$ and $Q' _2$ are (imposing spherical symmetry) given by integrals of the form

\begin{eqnarray}
Q' _1 = \frac{4 \pi}{9 G} \int \limits _{0} ^{R_{\rm core}} \frac{r^{7} \rho(r)}{M(r)} \D r \comequa \\
Q' _2 = \frac{4 \pi}{9 G} \int \limits _{R_{\rm core}} ^{R_{\rm sec}} \frac{r^{7} \rho(r)}{M(r)} \D r \comequa
\end{eqnarray}

with the secondary's radial density profile $\rho(r)$ and $M(r)$ being the total mass inside a radius $r$. Solving Eq.~\ref{eq:omega2equa} for $\Delta \Omega _2$ and using

\begin{equation}
\label{eq:deltaE}
\Delta E = ( \Omega _2 - \Omega _1 ) \cdot I_2 \Delta \Omega _2 + \frac{1}{2} \left[ \frac{1}{I_1} + \frac{1}{I_2} \right] ( I_2 \, \Delta \Omega _2 ) ^{2} 
\end{equation}

with the moment of inertia 

\begin{eqnarray}
I_1 = \frac{8 \pi}{3} \int \limits _{0} ^{R_{\rm core}} r^{4} \rho (r) \D r \comequa \\ 
I_2 = \frac{8 \pi}{3} \int \limits _{R_{\rm core}} ^{R_{\rm sec}} r^{4} \rho (r) \D r \comequa
\end{eqnarray}

gives the total amount of energy needed to perform the angular momentum transfer. 

The total number of parameters can be reduced by imposing angular momentum conservation, i.e. a lossless exchange 

\begin{equation}
I_1 \Delta \Omega _1 + I_2 \Delta \Omega _2 = 0 \fsequa
\end{equation}

Following \citet{Brinkworth2006}, the minimum energy required to drive Applegate's mechanism is achieved assuming no initial differential rotation or

\begin{equation}
\Omega _1 = \Omega _2 \fsequa
\end{equation}

Altogether, we can now solve Eq.~\ref{eq:omega2equa} for $\Delta \Omega _2$ and find

\begin{equation}
\label{eq:solutionOmega2}
( \Delta \Omega _2) = -\Omega_2 \, \frac{\beta}{\alpha} \pm \sqrt{ \, \left[ \Omega _2 \, \frac{\beta}{\alpha} \right]^{2} + \frac{\Delta Q}{\alpha} } \comequa
\end{equation}

where we defined

\begin{equation}
\alpha := \gamma ^{2} Q' _1 + Q' _2 \comequa ~~ \beta := - \gamma Q' _1 + Q' _2 \comequa ~~ \gamma := \frac{I_2}{I_1} 
\end{equation}

for convenience and as a preparation for the next sections. 

\subsection{The case of a constant density profile}\label{sec:constant}
As a zero-order approximation, the framework introduced by \citet{Brinkworth2006} can be evaluated assuming a constant density throughout the star. While this is clearly an approximation, it already gives rise to some phenomenological scaling relations to illustrate how the required energy for the Applegate mechanism depends on the binary separation and the mass of the secondary. Employing this approximation, the density in the secondary is now given as
\begin{equation}
\rho (r) = \bar{\rho} = \frac{3 M_{\rm sec}}{4 \pi R _{\rm sec} ^{3}} \fsequa
\end{equation}

Given this and defining $ \xi = R_{\rm sec}/R_{\rm core} $, we can calculate explicit expressions for the coefficients in eq.~\ref{eq:solutionOmega2}:

\begin{align*}
\alpha = \frac{R_{\rm core}^{5}}{15 G} \cdot ( \xi ^{10} - \xi ^{5} ) \comequa ~~ \beta = 0 \comequa ~~ \gamma = \xi ^{5} - 1 \fsequa
\end{align*}

In this context, $\Delta \Omega _2$ is given as

\begin{equation}
\Delta \Omega _2 = \pm \sqrt{\frac{15G \, \Delta Q}{R_{\rm core} ^{5} \, [\xi ^{10} - \xi ^{5}]}} \fsequa
\end{equation}

Inserting this into eq.~\ref{eq:deltaE} eliminates both the dependence on $R_{\rm core}$ and $\Omega _2$ and finally yields

\begin{equation}
\centering
\Delta E = \frac{G}{3} \cdot \left( \frac{\Delta P}{P_{\rm bin}} \right) \cdot \frac{a_{\rm bin}^{2} \, M_{\rm sec}^{2}}{R_{\rm sec}^{3}} \comequa
\end{equation}

or in a more practical form

\begin{equation}
\label{eq:delta_e_simple}
\Delta E \simeq 1.3 \cdot 10^{48} \erg \, \left( \frac{\Delta P}{P_{\rm bin}} \right) \, \left( \frac{a_{\rm bin}}{\Rsol} \right)^{2} \, \left( \frac{M_{\rm sec}}{\Msol} \right)^{2} \, \left( \frac{R_{\rm sec}}{\Rsol} \right)^{-3},
\end{equation}

which can also be expressed in terms of a relative Applegate threshold energy via division by the energy provided over one modulation period by the secondary:

\begin{equation}
\frac{\Delta E}{E_{sec}} \simeq 1.1 \cdot 10^{7} \, \left( \frac{\Delta P}{P_{\rm bin}} \right) \, \left( \frac{a_{\rm bin}}{\Rsol} \right)^{2} \, \left( \frac{M_{\rm sec}}{\Msol} \right)^{2} \, \left( \frac{R_{\rm sec}}{\Rsol} \right)^{-3} \, \left( \frac{P_{\rm mod}}{\yr} \right)^{-1} \, \left( \frac{L_{\rm sec}}{\Lsol} \right)^{-1}.
\end{equation} 

The main goal of this calculations is a simple model to give rough order of magnitude estimates as well as the basic scaling properties of the Applegate mechanism. It is interesting to note that this zero-order estimate is independent of the angular velocity of the star, and does not require an assumption of stellar rotation. The results of this estimate along with those for the improved models are given in Table~\ref{tab:results}. In the two-zone model presented in the next subsection, the required energy will include a dependence on the rotation rate, which can be estimated from the assumption of tidal locking.

\subsection{An analytical two-zone model}\label{twozone}

\subsubsection{Derivation}

While the model above already provides an interesting scaling relation including the dependence on binary separation and the mass of the secondary, it is instructive to further understand the dependence on the stellar structure and the rotation of the secondary. For this purpose, we consider now a stellar density profile with a density $\rho_1$ in the core and $\rho_2$ in the shell. We denote the size of the core as $R_{\rm core}$. The density profile is then given as

\begin{equation}
\rho(r) = \begin{cases} 
\rho_1 & 0 \leq r \leq R_{\rm core}, \\
\rho_2 & R_{\rm core} < r \leq R_{\rm sec},
\end{cases}
\end{equation} 

and the main parameters can be summarized as

\begin{equation}
\lambda = \rho _2 / \rho _1 ~~~,~~~ \xi = R_{\rm star} / R_{\rm core}.
\end{equation}

We can calculate that

\begin{align*}
\alpha = \frac{\lambda R_{\rm core}^{5}}{15 G} \cdot \left( \lambda [\xi ^{5} -1]^{2} + f(\xi, \lambda) \right), \\
\beta = \frac{\lambda R_{\rm core}^{5}}{15 G} \cdot \left( f(\xi,\lambda) - \gamma \right), \\
\gamma = \lambda \left[ \xi ^{5} - 1 \right], 
\end{align*}

where the function $f$ is given by

\begin{equation}
f(\xi, \lambda) = \int \limits ^{\xi} _{1} \frac{5 \cdot x^{7}}{1 - \lambda + \lambda x^{3}} \D x.
\end{equation}

The two solutions for the angular velocity change are given as

\begin{equation}
\Delta \Omega _2 = -\Omega_2 \cdot \frac{\beta}{\alpha} \cdot \left( 1 \pm \sqrt{1 + \frac{\alpha \Delta Q}{\beta^{2} \Omega^{2} _2 }} \, \right)  
\end{equation}

with

\begin{equation}
\Delta Q = - \frac{a_{\rm bin}^{2} M_{\rm sec}}{9} \frac{\Delta P}{P_{\rm bin}}.
\end{equation}

We assume now that the star is tidally locked, implying that the orbital period of the secondary is given as the orbital period of the binary system, i.e. $\Omega _2 = 2 \pi / P_{\rm bin}$. Explicitly, we then have

\begin{equation}\label{eq:del_om2_tzm}
\Delta \Omega _2 = - \frac{2 \pi}{P_{\rm bin}} \frac{f-\gamma}{\gamma^{2}/\lambda + f} \left( 1 \pm \sqrt{1 - G k_2 \, \frac{a_{\rm bin}^{2} M_{\rm sec} P_{bin}^{2}}{R_{\rm sec}^{5}} \frac{\Delta P}{P_{\rm bin}}} \right)
\end{equation}

with the coefficient

\begin{equation}\label{eq:tzm_k2}
k_2 = \frac{15}{36 \pi^{2}} \, \frac{\xi^{5}}{\lambda} \, \frac{(\gamma^{2}/\lambda + f)}{(f-\gamma)^{2}} \fsequa
\end{equation}

The energy required to drive the angular momentum exchange is

\begin{equation}
\Delta E = \frac{1}{2} (\gamma +1) (\xi^{5}-1) \frac{8 \pi \rho _2}{15} R_{\rm core}^{5} \Delta \Omega _2 ^{2},
\end{equation}

which together with the identity

\begin{equation}
\frac{4 \pi}{3} \bar{\rho} R_{\rm star}^{3} = \frac{4 \pi}{3} \rho_1 R_{\rm core}^{3} + \frac{4 \pi}{3} \rho_2 ( R_{\rm star}^{3} - R_{\rm core}^{3} ).
\end{equation}

results in

\begin{equation}
\label{eq:tzm_energy}
\Delta E = \frac{1}{5} \frac{(\gamma+1)(\xi^{3}-\xi^{-2})}{1+\lambda(\xi^{3}-1)} \lambda M_{\rm star} R_{\rm star}^{2} \Delta \Omega^{2} _2,
\end{equation}

or with Eq.~\ref{eq:del_om2_tzm} 

\begin{equation}
\label{eq:tzm_energy_full}
\Delta E = k_1 \cdot \frac{M_{\rm sec} R_{sec}^{2}}{P_{\rm bin}^{2}} \cdot \left( 1 \pm \sqrt{1 - k_2 \, G \, \frac{a_{\rm bin}^{2} M_{\rm sec} P_{bin}^{2}}{R_{\rm sec}^{5}} \frac{\Delta P}{P_{\rm bin}}} \right)^{2},
\end{equation}

where we defined the coefficient

\begin{equation}\label{eq:tzm_k1}
k_1 = \frac{4 \pi^{2}}{5} \, \frac{\lambda (\gamma+1) (\xi^{3}-\xi^{-2})}{1+\lambda (\xi^{3}-1)} \, \frac{(f-\gamma)^{2}}{(\gamma^{2}/ \lambda +f)^{2}} \fsequa
\end{equation}

According to the last section, we can relate eq.~\ref{eq:tzm_energy_full} to the energy provided over one modulation period:

\begin{equation}
\frac{\Delta E}{E_{sec}} = k_1 \cdot \frac{M_{\rm sec} R_{sec}^{2}}{P_{\rm bin}^{2} P_{mod} L_{sec}} \cdot \left( 1 \pm \sqrt{1 - k_2 \, G \, \frac{a_{\rm bin}^{2} M_{\rm sec} P_{bin}^{2}}{R_{\rm sec}^{5}} \frac{\Delta P}{P_{\rm bin}}} \right)^{2}.
\end{equation}

For low-mass stars, typical values are $\xi \sim 4/3$ and $\lambda \sim 1/100$ (see sec.~\ref{sec:results}). As the inverse zone transition parameter $\xi$ shows just negligible variations for typical PCEB systems, we can fix it to $\xi = 4/3$ to find the zone density contrast parameter $\lambda$ which is most compatible with the systems examined in this paper (see tab.~\ref{tab:secondarys}). We set $\lambda = 0.00960$, implying $f \sim 5.57$. Given that, we can numerically evaluate the coefficients yielding

\begin{equation}
k_1 = 0.133 ~~~,~~~ k_2 = 3.42 \fsequa
\end{equation}

\subsubsection{A critical condition}

In contrast to the constant density approximation, the two-zone model incorporates all essential physics involved in the Applegate process. In particular, it accounts for the orbital period of the binary resulting in two different energy branches just as in the full treatment (see sec.~\ref{sec:brinkworth}). Here, the lower energy branch corresponds to the negative solution. Another implication results from the term inside the root: Restricting us to real-valued solutions raises the critical condition 

\begin{equation}
\label{eq:tzm_condition}
k_2 \, G \, \frac{a_{\rm bin}^{2} P^{2}_{\rm bin} M_{\rm sec}}{R_{\rm sec}^{5}} \frac{\Delta P}{P_{\rm bin}} := A \leq 1 \fsequa
\end{equation}  

Mathematically, systems which do not satisfy eq.~\ref{eq:tzm_condition} (hereafter, we refer to the left-hand side as the Applegate parameter $A$) cannot drive the observed period change independent of energetic arguments as no real-valued solution exists. On the other hand, one can show that in the case of a critical system for which the Applegate parameter is unity, the two-zone model and the constant density model converge. In such a system, the energy to drive the Applegate process is given by

\begin{equation}
\Delta E = k_1 \cdot \frac{M_{\rm sec} R_{sec}^{2}}{P_{\rm bin}^{2}} \fsequa
\end{equation} 

Substituting eq.~\ref{eq:tzm_condition} for the binary period leads to

\begin{equation}
\Delta E = k_1 \, k_2 \, G \, \cdot \left( \frac{\Delta P}{P_{bin}} \right) \cdot \frac{a_{\rm bin}^{2} \, M_{\rm sec}^{2}}{R_{\rm sec}^{3}} \fsequa
\end{equation}

In the case of constant density, $\lambda=1$ which means that $f=\xi^{5}-1 = \gamma$ and we end up at $k_1 \cdot k_2 = 1/3$, proving that the two-zone model and the constant density model give identical Applegate energies for critical systems. Using this, we can understand the physical meaning of critical systems by looking into the angular momentum budget of the star and its rotational state. For a critical system, the change of the outer shell's angular velocity (cf. eq.~\ref{eq:del_om2_tzm}) is given by

\begin{equation}
\Delta \Omega _2 = -\frac{2 \pi}{P_{bin}} \cdot \frac{f - \gamma}{\gamma^{2}/\lambda+f} \fsequa
\end{equation} 

Now, let $\lambda \rightarrow 1$ which implies $f \rightarrow \gamma$ because for critical systems the two-zone model and the constant density model converge as we showed above. Non-zero solutions impose $\xi \rightarrow 1$ and we arrive at

\begin{equation}
\Delta \Omega _2 = \lim\limits_{\xi \rightarrow 1}{-\frac{2 \pi}{P_{bin}} \cdot \frac{\xi^{5}-1 - \xi^{5}+1}{\xi^{10}-\xi^{5}+\xi^{5}-1}} = -\frac{2 \pi}{P_{bin}} 
\end{equation}

proving that the constant density model is the limit of a two-zone calculation for the extreme case that the outer shell and its angular velocity vanish.

\subsubsection{Quality of the approximation}

The quality of the analytical two-zone model becomes clear from Fig.~\ref{fig:tzm_test} where we compare the estimates as calculated with the two-zone model with the full calculations described in the next subsection. The full model considers a white dwarf primary with $0.5~\Msol$ accompanied by a fairly evolved secondary star with $t \sim 5~\Gyr$, while the two-zone model employs $\lambda=0.0096$ and $\xi =4/3$, consistent with the typical structure of a low-mass star. In both calculations, we assume a relative period change of $\Delta P / P_{bin} = 10^{-7}$ with a modulation period of $P_{\rm mod} = 14~\yr$, corresponding to a Jupiter-like planet with mass $\sim 3~\Mjup$ and semi-major axis $\sim 5~\au$ and we investigate the results for varying Applegate parameters $A$. As one can see, the typical deviation of the required energy for the Applegate mechanism corresponds to less than $10\%$ between the two models, provided that the Applegate condition (Eq.~\ref{eq:tzm_condition}) is satisfied with a left-hand side much smaller than unity. We adopt $A \lesssim 0.5$ as a typical fiducial limit.

\bef
\resizebox{\hsize}{!}{\includegraphics{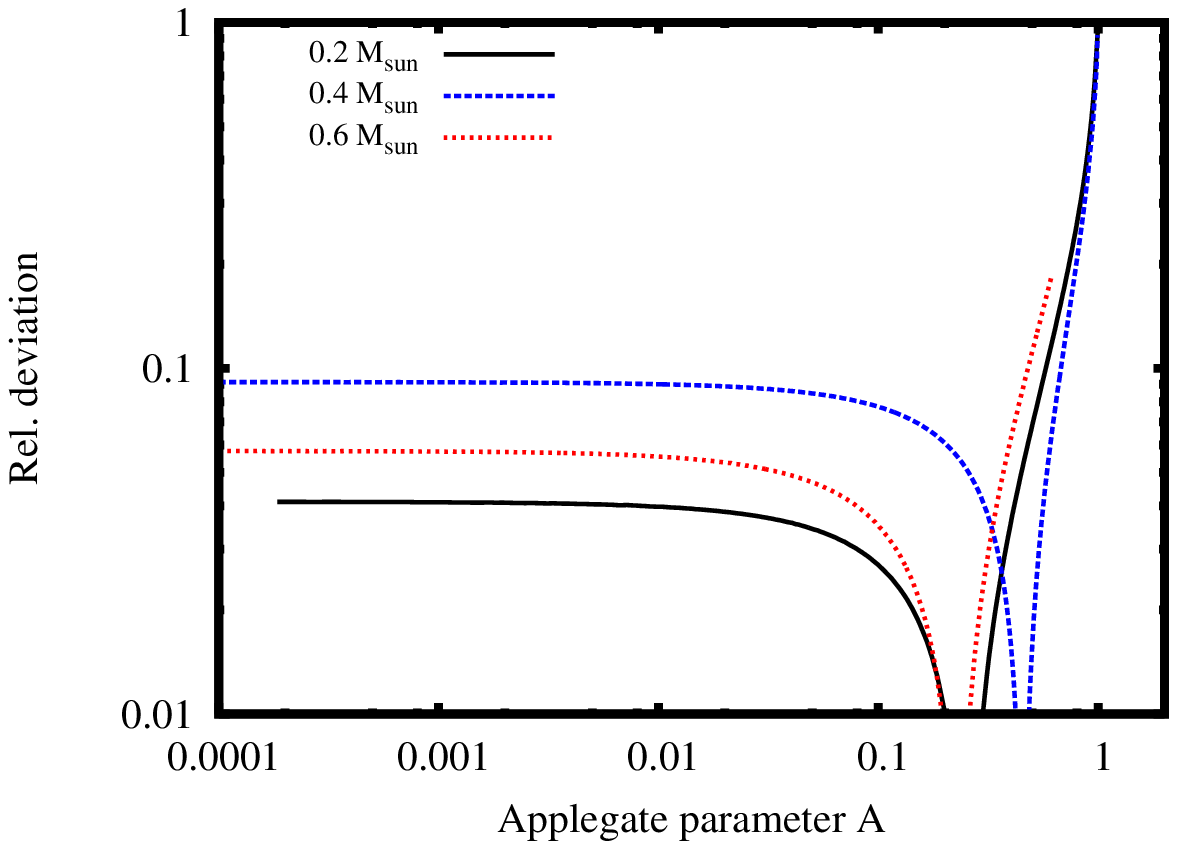}}
\caption{The relative difference between the two-zone model described in section \ref{twozone} with the full model considering a realistic stellar density profile as outlined in section \ref{sec:recipe}. The model assumes a white dwarf mass of $0.5$~M$_\odot$ and an age of the secondary of $5$~Gyr. As parameters in the two-zone model, we employ $\lambda=0.0096$ and $\xi=4/3$. The relative deviation of the results is shown as a function of the Applegate parameter $A$ and for different secondary masses. A good approximation requires $A \lesssim 0.5$.}
\label{fig:tzm_test}
\eef

Finally, in Fig.~\ref{fig:comp_models} we compare the computational results of all three models described in the last two sections, namely the constant density approximation, the two-zone model as well as the full model for the same general system configuration as in Fig.\ref{fig:tzm_test} and varying binary separation. Here, we fixed the secondary mass to $M_{sec}=0.3~\Msol$ allowing us to compare the models for varying Applegate parameter. Over almost the entire parameter range, the two-zone model precisely resembles the full calculations with absolute deviations of less than $10\%$ while the constant density model lies off by several orders of magnitude. Both the full calculations and the two-zone model fail beyond $A \sim 1$ (corresponding to $a_{bin} \sim 3.5~\Rsol$), which is exactly predicted by eq.~\ref{eq:tzm_condition}.

We therefore conclude that the two-zone model provides a valuable and sufficient approximation to estimate the required energy for the case of PCEBs with main sequence low-mass companions in the range of $0.1~\Msol$ to $0.6~\Msol$ that satisfy the Applegate condition eq.~\ref{eq:tzm_condition}. Nevertheless we will evaluate our main results with the more detailed numerical framework which includes full stellar density profiles and varying core radii.

\bef
\resizebox{\hsize}{!}{\includegraphics{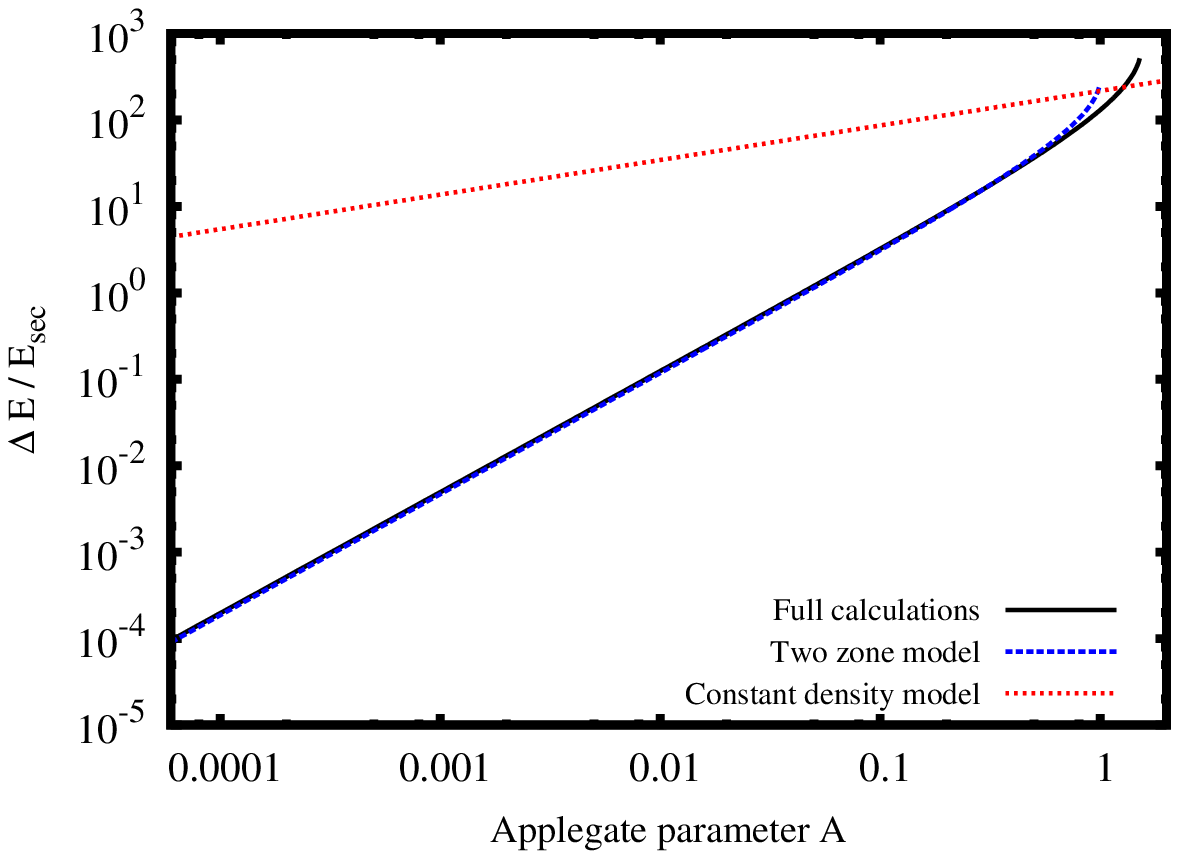}}
\caption{In this figure we compare the two analytical models, namely the constant density and the two-zone models as described in sec.~\ref{sec:constant} and sec.~\ref{twozone}, with the full calculations as described in sec.~\ref{sec:recipe} in terms of the predicted relative Applegate threshold energy. We consider the same general system configuration as in Fig.~\ref{fig:tzm_test} and described in sec.~\ref{twozone} but fixed the secondary mass to $M_{sec} = 0.3~\Msol$ enabling us to plot the calculated energies as functions of the Applegate parameter $A$ (see eq.~\ref{eq:tzm_condition}). The analytical two-zone model closely follows the full calculations with typical deviations of less than $10\%$ while the simple constant density model overestimates the energy threshold by several magnitudes except for the critical region close to unity.}
\label{fig:comp_models}
\eef

\section{Full calculations with a stellar density profile}
\subsection{Recipe}
\label{sec:recipe}

In addition to the constant density profile and the two-zone model, we present here the framework to derive the required energy to drive Applegate's mechanism based on detailed and realistic density profiles. The latter requires a numerical solutions for the coefficients involved in Eq.~\ref{eq:solutionOmega2}. In this section, we will describe our general framework and apply it to NN Ser. The results for the additional systems will be given in section~\ref{sec:results}.\\

From Eq.~\ref{eq:solutionOmega2}, we expect that the effective change of the shell's angular velocity $\Delta \Omega _2$ is an explicit function of its initial angular velocity $\Omega _2$ and an implicit function of the core radius $R_{core}$ via the moments of inertia $I_{1,2}$ and the $Q'_{1,2}$ coefficients. Our procedure is thus as follows: 

First we calculate $I_{1,2}$ and $Q'_{1,2}$ for a given core radius $R_{\rm core}$ utilizing a radial density profile provided by \textit{Evolve ZAMS}\footnote{\resizebox{\hsize}{!}{Webpage Evolve ZAMS: http://www.astro.wisc.edu/\~townsend/static.php?ref=ez-web}} \citep{Paxton2004}, but re-scaled to the secondary's radius and normalized to its mass as inferred by \citet{Parsons2010NNSer} to resemble the measurements. We adopt an age value of $\sim 2~\Gyr$ which is roughly the main sequence lifetime of the $2~\Msol$ WD progenitor \citep[see][]{Beuermann2010}. Unless stated otherwise, we will in the following assume a solar metallicity.

Based on that, we can calculate the two solutions for $\Delta \Omega _2$ for a given initial rotation $\Omega _2$, named $\Delta \Omega _2 ^{+}$ for the positive sign and $\Delta \Omega _2 ^{-}$ respectively. These two solutions finally give two different corresponding energies $\Delta E ^{+}$ and $\Delta E ^{-}$.

According to \citet{Haefner2004}, the NN Ser system is tidally locked, constraining the inital angular velocities to

\begin{equation}
\Omega _1 = \Omega _2 = \frac{2 \pi}{P_{\rm bin}} := \Omega _{\rm bin} \comequa
\end{equation}

with $P_{\rm bin}$ the orbital period of the binary \citep[see, e.g.,][]{Brinkworth2006,Lanza2006}. Given the  condition of tidal locking, the only remaining parameter is the core radius $R_{\rm core}$. We do not restrict it to equal the nuclear burning zone as typical Red Dwarfs are fully convective \citep[see, e.g.,][]{Engle2011}, but rather explore the minimum energy that can be obtained depending on the core radius. The latter is parametrized as a fraction $\delta$ of the core radius. The relative period change $\Delta P / P_{\rm bin}$ of the binary is calculated assuming sinusoidal perturbance by planets on circular orbits via

\begin{equation}
\frac{\Delta P}{P_{\rm bin}} = 4 \pi \frac{K}{P_{\rm mod}}
\end{equation}  

with semi-amplitude $K$ and modulation period $P_{\rm mod}$ \citep[see, e.g.,][]{Applegate1992, Gozdziewski2015}, neglecting small variations due to orbital eccentricity in both the binary or the planet.

\bef
\resizebox{\hsize}{!}{\includegraphics{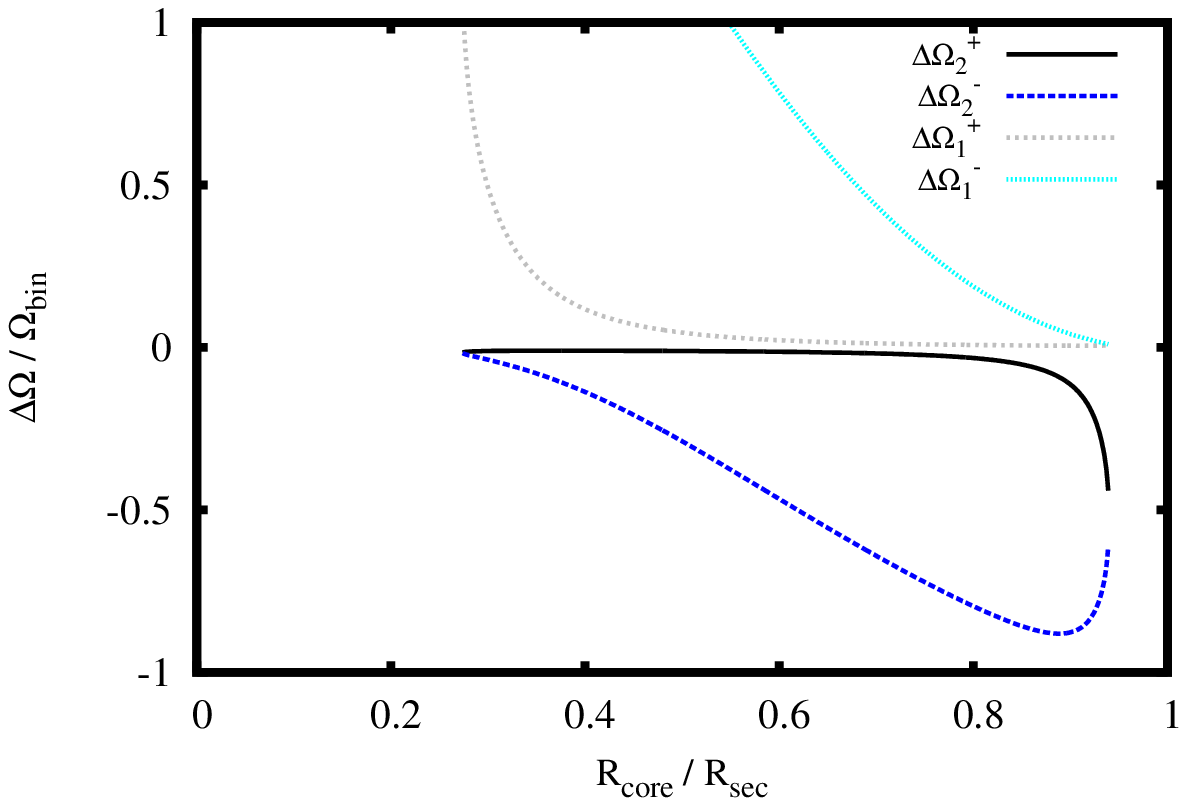}}
\caption{The angular velocity change normalized by the angular velocity of the binary both in the core ($\Delta \Omega_1$) and the shell ($\Delta \Omega_2$) for our reference system NN Ser as a function of the size of the core. For both quantities, there are two solutions, denoted with + and -. The calculation is based on the full numerical model described in section~\ref{sec:recipe}.}
\label{fig:nnser_omega}
\eef

\bef
\resizebox{\hsize}{!}{\includegraphics{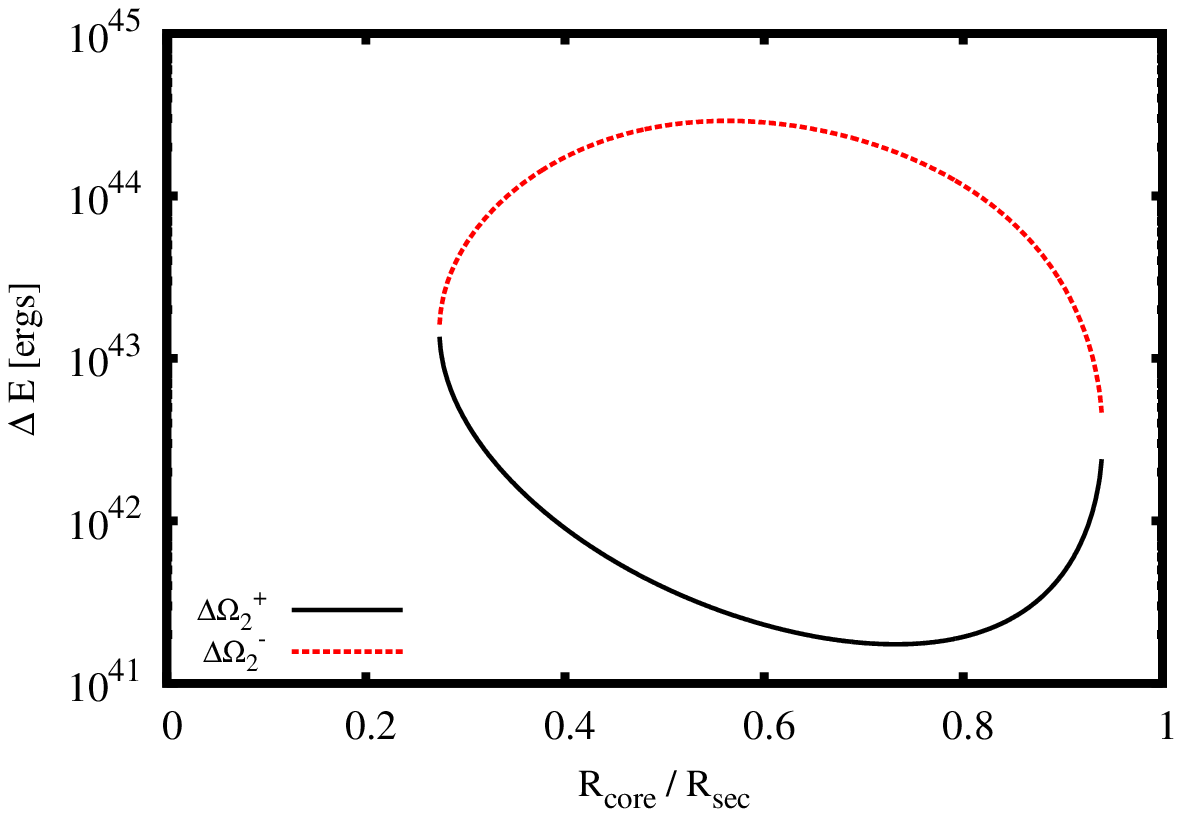}}
\caption{The required energy to drive the eclipsing time observations for the solutions denoted with + and - in Fig.~\ref{fig:nnser_omega}. The calculation is based on the full numerical model described in section~\ref{sec:recipe}.}
\label{fig:nnser_energy}
\eef

As one can see from Fig.~\ref{fig:nnser_omega}, we have two possible solutions corresponding to two different angular momentum transfer modes that are compatible with the boundary conditions. Both modes correspond to different energies. We plot them in Fig.~\ref{fig:nnser_energy} as a function of the fractional core radius $\delta := r_{\rm core} / r_{\rm sec}$, showing that the two branches converge as $\delta$ runs to either $0.3$ or $0.9$, but typically feature a prominent spread of several magnitudes in between. Outside this parameter space, only complex-valued solutions exist. The minimum energy for the modified Applegate mechanism (hereafter $\Delta E_{\rm min}$) is achieved at a fractional core radius of $\delta _{\rm min} \sim 0.75$ for the lower branch.\\

Finally, we compare the energy needed to drive the Applegate mechanism over one modulation period $P_{\rm mod}$ with the total energy generated by the secondary, i.e.

\begin{equation}
E_{\rm sec} = P_{\rm mod} \cdot L_{\rm sec} \fsequa
\end{equation}

In order to be viable, we require 

\begin{equation}
\Delta E_{\rm min} < E_{\rm sec} \comequa
\end{equation} 

 knowing that a more realistic condition would require $\Delta E_{\rm min} \ll E_{\rm sec}$, as the energy required for the quadrupole moment oscillations should correspond to a minor fraction of the available energy produced by the star. 

In Tables~\ref{tab:secondarys} and \ref{tab:planets}, we summarize the basic properties of all systems investigated in this paper. Our sample is based on the compilation of PCEBs  with potential planets described by \citet{Zorotovic2013}, as well as four RS~CVn binaries RU~Cnc, AW~Her \citep{Tian2009}, HR~1099 \citep{Fekel1983, Garcia2003, Frasca2004} and SZ~Psc \citep{Jakate1976, Popper1988, Wang2010}, and the RR-Lyr type binary BX~Dra \citep{Park2013}. 

\subsection{How does the age affect our results?}\label{sec:time}

Because of their low mass and luminosity, typical Red Dwarfs of $\sim 0.1 \Msol$ have main sequence lifetimes of several $100 \Gyr$ \citep[see, e.g.,][]{Engle2011}. On timescales of tens of $\Gyr$, their fundamental properties and internal structures remain virtually constant.\\

Even for more massive dwarfs such as in \textit{QS Vir} with a mass of $0.43~\Msol$, the radial density profile calculated with \textit{Evolve ZAMS} shows little variation even for the two extreme cases of $t \sim 1~\Gyr$ and $t \sim 14~\Gyr$: while its radius increased by $\sim 1~\%$, the core density increased by slightly more than $\sim 10~\%$, concentrating more mass in the center (see Fig.~\ref{fig:qsvir_density}). Calculating the relative energy threshold to drive the Applegate process as described in sec.~\ref{sec:recipe} assuming $\Delta P / P_{bin} = 10^{-6}$, we find:

\begin{itemize}
\item $t = 1~\Gyr$: $\Delta E_{min}/E_{sec} = 0.615$ at $\delta = 0.729$
\item $t = 14~\Gyr$: $\Delta E_{min}/E_{sec} = 0.692$ at $\delta = 0.713$
\end{itemize}  

As both results differ by just $\sim 10 \%$, we conclude that the age of the system is a higher-order effect for the typical systems examined in this paper. We therefore adopt a canonical value of $t = 5~\Gyr$ if no age estimates are given in the literature. For the binaries with evolved components, we estimate the age with \textit{Evolve ZAMS} by solving for the measured stellar radii.   

\bef
\resizebox{\hsize}{!}{\includegraphics{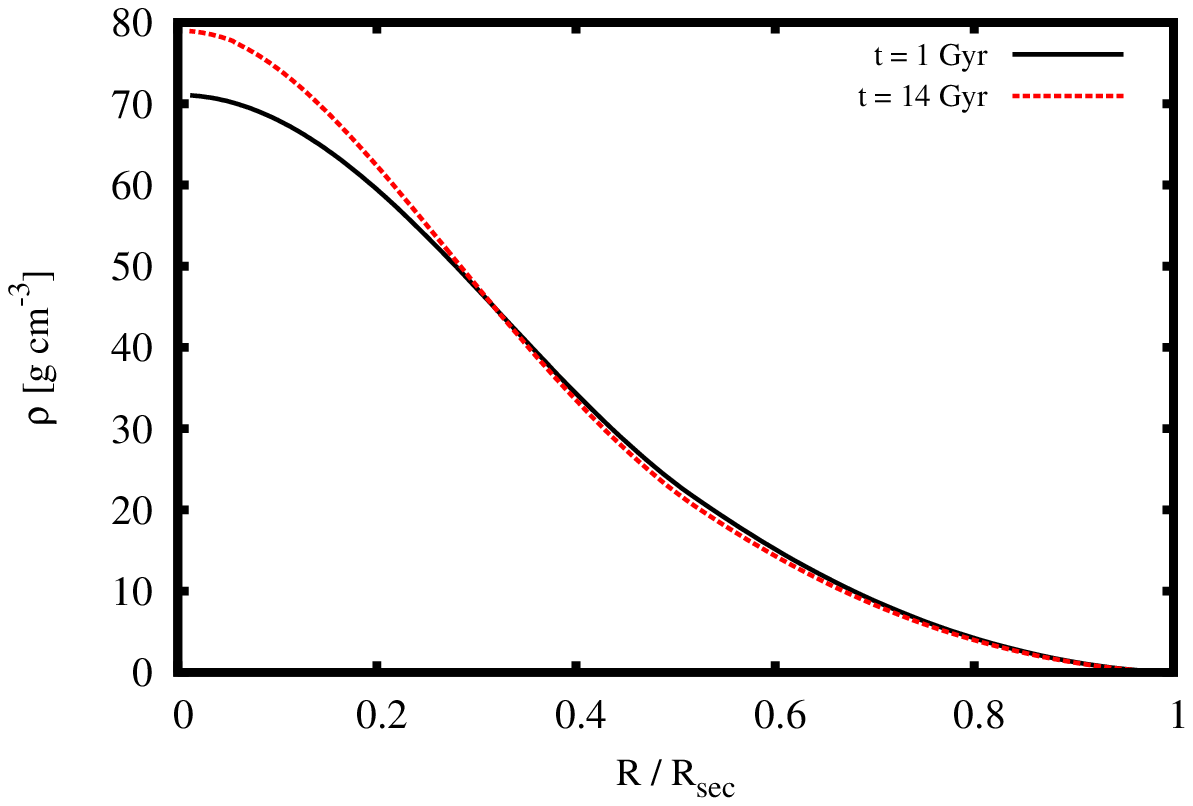}}
\caption{The density profile of a red dwarf with $0.1$~M$_\odot$ for a lifetime of $1$~Gyr and $14$~Gyrs. The variation in the core density due to the different lifetime corresponds to only about $10\%$. The calculation assumes a solar metallicity.}
\label{fig:qsvir_density}
\eef

\subsection{Dependence on metallicity variations}\label{sec:metallicity}
Here we explore how much the model results can be affected by uncertainties in the metallicity variation. Using \textit{Evolve ZAMS}, we have calculated the stellar density profile for the secondary in NN~Ser for metallicities between $10^{-4}$ and $3\times10^{-2}$, assuming a generic age of $5$~Gyrs. The resulting density profile is given in Fig.~\ref{fig:qsvir_metallicity}, showing a maximum variation in the core density of about $20\%$. 

We further summarize the resulting changes in the stellar luminosity, radius and surface temperature in Table~\ref{tab:metallicity}. While the maximum variations in radius and surface temperature correspond  to about $20\%$, the variation in the luminosity corresponds to a factor of $2$ between the extreme cases. Considering that the metallicity is varied by more than two orders of magnitude, the latter still corresponds to a minor uncertainty.

\begin{table}
\caption{Fundamental calculated secondary parameters for varying metallicity values.}
\begin{center}
\begin{tabular}{c|c|c|c}
\hline
Metallicity 	& Luminosity/$\Lsol$	& Radius/$\Rsol$ 		& Surf.temp./$\K$ 	\\
\hline
Z=0.0001		& 	0.0444 				&	0.377	& 4,320 	\\
Z=0.004			& 	0.0325 				&	0.398	& 3,890 	\\
Z=0.03			& 	0.0255 				&	0.401	& 3,650 	\\
\hline
\end{tabular}
\label{tab:metallicity}
\end{center}
\end{table}

\bef
\resizebox{\hsize}{!}{\includegraphics{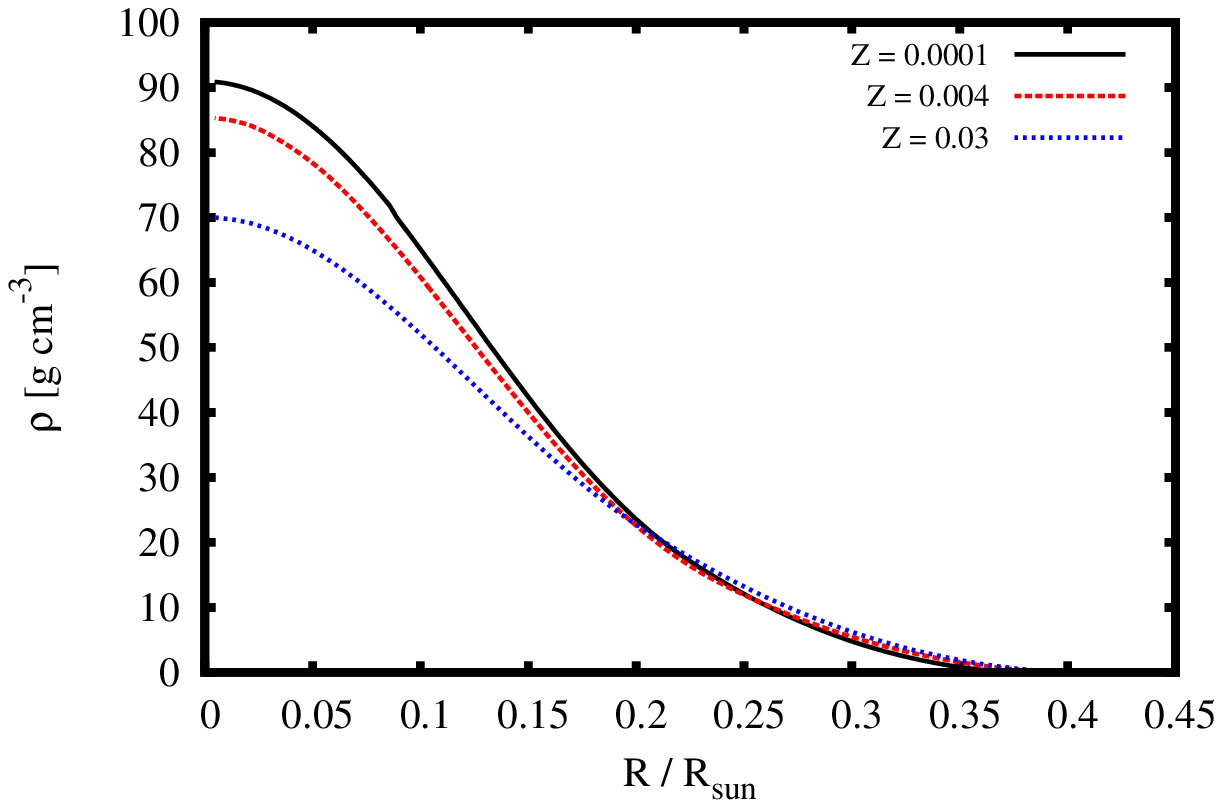}}
\caption{The density profile of a red dwarf with $0.1$~M$_\odot$ for metallicities between $10^{-4}$ and $3\times10^{-2}$. The maximum variation in the core density corresponds to about $20\%$. The calculation assumes a generic age of $5$~Gyrs.}
\label{fig:qsvir_metallicity}
\eef

Using the calculated stellar density profiles as input, we can now use the formalism developed in section~\ref{sec:recipe} to determine the required energy to drive the eclipsing time variations $\Delta E_{\rm min}$ and compare them to the energy $E_{\rm sec}$ produced by the secondary within one modulation period. For the metallicities investigated here, we find the following results:

\begin{itemize}
\item $Z = 0.0001$: $\Delta E_{min}/E_{sec} = 0.593$
\item $Z = 0.004$: $\Delta E_{min}/E_{sec} = 0.502$
\item $Z = 0.03$: $\Delta E_{min}/E_{sec} = 0.649$
\end{itemize}  

Again, the scatter in the ratio $\Delta E_{\rm min}/E_{\rm sec}$ is in the $10\%$-$20\%$ level even for the large range of metallicities considered here. While the latter contributes to the overall uncertainty, an incorrect metallicity estimate cannot significantly affect the question whether the Applegate mechanism is feasible or not.

\section{Results}
\label{sec:results}
In this section, we will apply the framework from the previous section (including the detailed model presented in section~\ref{sec:recipe} to investigate how the feasibility of the Applegate model depends on the properties of the binary system). We will further apply it to the sample presented in section~\ref{sec:systems} to assess for which of these systems the eclipsing time observations can be explained with the Applegate mechanism.

\subsection{Parameter study}\label{sec:param_study}

In the following, we consider a close binary system with varying separation consisting of a $0.5~\Msol$ White Dwarf and a Red Dwarf companion with different masses in the range of $0.15~\Msol$ to $0.6~\Msol$. We assume a fixed relative period variation $\Delta P / P_{bin} = 10^{-7}$ with a modulation period of $P_{\rm mod} = 14~\yr$, corresponding to a Jupiter-like planet with mass $\sim 3~\Mjup$ and semi-major axis $\sim 5~\au$.

Using the model presented in section~\ref{sec:recipe}, we determine the required energy to drive the Applegate mechanism as a function of binary separation for different masses of the secondary. The results of this calculation are presented in Fig~\ref{fig:param_study}. For all secondary masses, the Applegate threshold energy scales positively with increasing binary separation, as a larger quadrupole moment variation must be generated with increasing binary separation for the same period variation to be produced. The ratio $E_{\rm min}/E_{\rm sec}$ decreases with increasing mass of the secondary, as more massive secondaries produce higher stellar luminosities and more energy that is potentially available to drive Applegate's mechanism. The latter implies that more massive secondaries are particularly well-suited to produce quadrupole moment variations, while it is more difficult for low-mass companions as observed in NN~Ser.

Qualitatively, we can understand the scaling behavior seen in Fig.~\ref{fig:param_study} utilizing the constant-density model. Normalized to the energy provided by the secondary over one planetary orbit, we have

\begin{equation}
\frac{\Delta E}{E_{\rm sec}} \propto \frac{\Delta P}{P_{bin}} \, a_{\rm bin}^{2} \, M_{\rm sec}^{2} \, R_{\rm sec}^{-3} \, L_{\rm sec}^{-1} \, P_{\rm plan}^{-1} \fsequa
\end{equation} 

From \citet{Demircan1991}, we adopt $R_{\rm sec} \propto M_{\rm sec}^{0.95}$ and $L_{\rm sec} \propto M_{\rm sec}^{2.6}$, while the relative period change and the planetary period are virtually constant. Combined we get

\begin{equation}
\frac{\Delta E}{E_{\rm sec}} \propto a_{\rm bin}^{2} \, M_{\rm sec}^{-3.45} \comequa
\end{equation}

resembling the fact that the relative Applegate threshold energy raises for increasing binary separation and decreases for increasing secondary mass. 

\bef
\resizebox{\hsize}{!}{\includegraphics{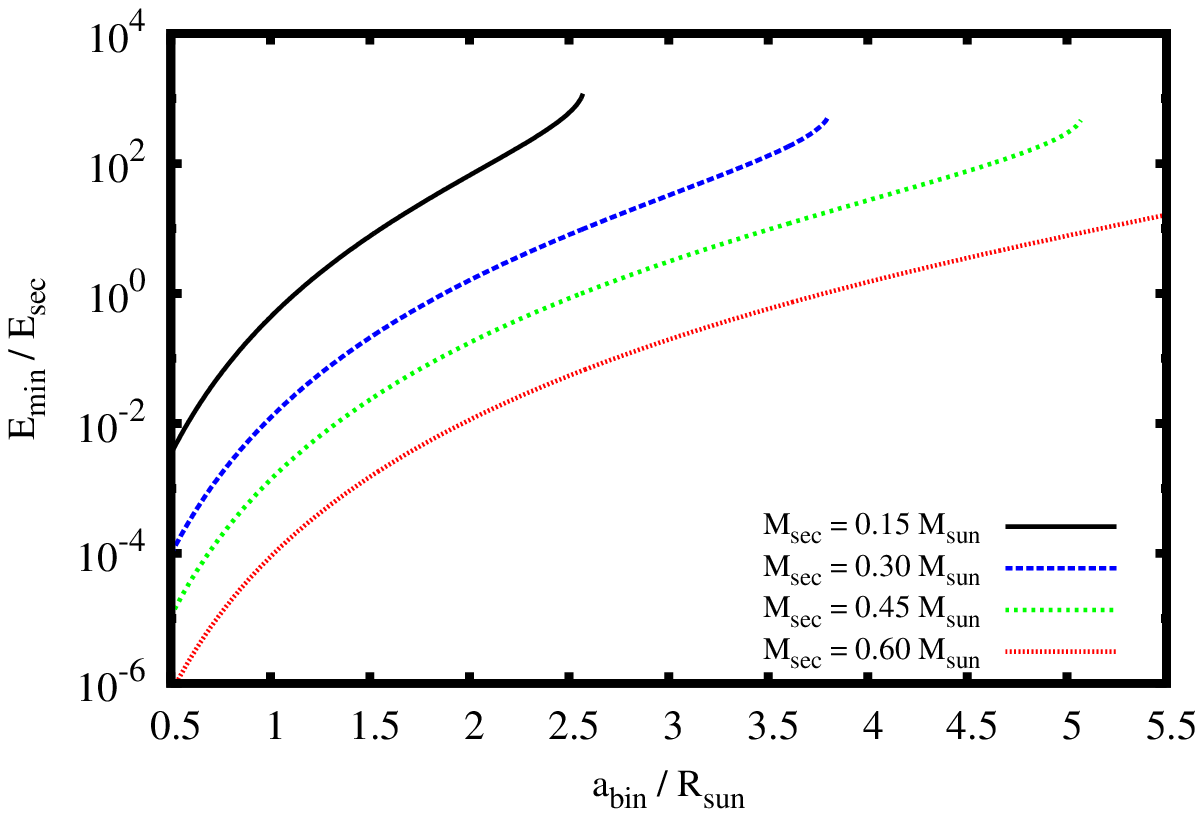}}
\caption{Relative Applegate energy calculated for varying secondary masses and binary separations according to sec.~\ref{sec:param_study}. Typical PCEBs with extremly low-mass secondarys do not provide enough energy to power the Applegate process or even do not satisfy the Applegate condition. Rather, typical Applegate systems are both relatively massive and extremely compact.}
\label{fig:param_study}
\eef

\subsection{Application to our sample}

We applied the calculations as described in section~\ref{sec:recipe} on the systems introduced and characterized in section~\ref{sec:systems}. In all our calculations, we assumed solar metallicity and rescaled and normalized the calculated radial density profile to resemble the (mean) observed mass and radius values. A summary of the main results is given in Table~\ref{tab:results}.

\begin{table*}
\caption{Summary of our calculations as described in sec.~\ref{sec:recipe}. The ratios $\Delta E_{\rm min}$/$E_{sec}$ denote the energy required to drive an Applegate mechanism of the observed magntiude over the available energy produced by the secondary star. The parameter $\delta_{\rm min}$ denotes the ratio of core radius to secondary star radius for which the minimum energy is obtained in the numerical model. The dash denotes imaginary values implying that no physical solution exists as discussed in sec.~\ref{twozone} and sec.~\ref{sec:recipe}.}
\begin{center}
\begin{tabular}{c|c|c|c|c|c|c|c}
\hline
System 	& $E_{\rm sec}$/$\erg$	&$\Delta E_{\rm min}$/$E_{sec}$ &$\Delta E$/$E_{\rm sec}$ 	& $\Delta E$/$E_{\rm sec}$ 		& $\Delta E$/$E_{\rm sec}$ 			& $\Delta E_{\rm min}$/$E_{\rm sec}$ & $\delta _{\rm min}$  \\
\hline
		&  						& \citet{Applegate1992} 		& \citet{Tian2009} 			& \multicolumn{4}{|c}{\centering This paper}						 \\
\hline
		&						& (see eq.~\ref{eq:applegate_energy})  & (see eq.~\ref{eq:tian}) & Const.dens. & Two-zone & \multicolumn{2}{|c}{Full model} \\
\hline
HS 0705+6700	& 	$2.2\cdot 10^{39}$ 		& 6.7	& 7.3	&	3,300	& 140 	& 140  	& 0.73 	\\
HW Vir			& 	$2.0\cdot 10^{40}$ 		& 4.9	& 6.0	&	720		& 108	& 104  	& 0.72 	\\
NN Ser			& 	$2.7\cdot 10^{39}$ 		& 3.2	& 3.3	&	1,100	& 64 	& 64  & 0.73 	\\
NSVS14256825	& 	$8.3\cdot 10^{38}$ 		& 5.3	& 5.4	&	3,200	& 101	& 102  	& 0.73 	\\
NY Vir			& 	$1.4\cdot 10^{39}$ 		& 5.5	& 5.6	&	2,800	& 106 	& 106 	& 0.73 	\\
HU Aqr			& 	$1.4\cdot 10^{40}$ 		& 0.10	& 0.10	&	240		& 1.9 	& 1.9  & 0.732 	\\
QS Vir			& 	$3.0\cdot 10^{40}$ 		& 0.039	& 0.040	&	170		& 0.71 & 0.77  & 0.71 	\\
RR Cae			& 	$5.2\cdot 10^{39}$ 		& 2.8	& 2.9	&	560		& 59	& 59  & 0.73 	\\
UZ For			& 	$4.1\cdot 10^{39}$ 		& 0.14	& 0.15	&	360		& 2.7	& 2.7  & 0.73 	\\
DP Leo			& 	$2.9\cdot 10^{39}$ 		& 0.021	& 0.021	&	150		& 0.38 & 0.38  & 0.74 	\\
V471 Tau		& 	$2.0\cdot 10^{42}$ 		& 0.014	& 0.014	&	12		& 0.26	& 0.26  & 0.84 	\\
\hline
RU Cnc			& 	$5.7\cdot 10^{43}$ 		& 0.074	& 0.076	&	1.7		& - 	& - 	& - \\
AW Her			& 	$8.5\cdot 10^{42}$ 		& 608	& 618	&	270		& - 	& - 	& - \\
HR 1099		    & 	$3.7\cdot 10^{43}$ 		& 0.21 	& 0.22	&	10		& - 	& 6.7  & 0.64 	\\
BX Dra			& 	$3.5\cdot 10^{43}$ 		& 0.00016& 0.00016&0.92		& 0.0029 & 0.056 & 0.52 \\
SZ Psc			& 	$9.9\cdot 10^{43}$ 		& 0.12	& 0.13	&	4.7		& - 	& 4.84 	& 0.61 \\
\hline
\end{tabular}
\label{tab:results}
\end{center}
\end{table*}


In total, we investigated 16 systems. Considering the numerical model based on realistic density profiles from section~\ref{sec:recipe}, the Applegate formalism can safely explain the period variations of four systems (QS Vir, DP Leo, V471 Tau, BX Dra). However, this may be an underestimate for QS~Vir, which contains variations in the O-C diagram much steeper than the average variation within one modulation time. For the other 12 systems, the relative threshold energy is greater than unity or no solution exists at all, implying that more than the total energy generated by the secondary is necessary to power the binary's period variations or the systems architecture is not capable of driving such a high level of period variation via the Applegate mechanism. We note that for four of these systems, in particular HU~Aqr, UZ~For, HR~1099 and SZ~Psc, the ratio $\Delta E/E_{\rm sec}$ is of order 1. Given the uncertainties regarding metallicity and age as discussed in the previous sections \ref{sec:time} and \ref{sec:metallicity}, these systems may still be able to drive an Applegate mechanism. However, for the remaining 8 systems, the ratio $\Delta E/E_{\rm sec}$ is considerably larger than $1$, implying that the observed eclipsing time variations cannot be explained by magnetic activitiy, particularly not in the wider, more massive and evolving RS CVn systems.

For comparison, we show the results from our constant density model, which tends to produce higher estimates of $\Delta E/E_{\rm sec}$, leading to an overestimate of the required energy. The two-zone model yields results close to the numerical values. For comparison, we also show the results adopting the original framework by \citet{Applegate1992} as presented by \citet{Parsons2010} as well as the fit by \citet{Tian2009}. Both cases tend to significantly underestimate the energy required to drive the eclipsing time variations, due to the thin-shell approximation and its inherent negligence of the core's backreaction.  We therefore emphasize that an assessment of the Applegate mechanism needs to be based at least on a two-zone model.



\subsection{Activity of the likely candidates}
In order to test the hypothesis of an Applegate mechanism, we have checked for the presence of magnetic activity those systems where the Applegate mechanism can be expected to produce period time variations of the observed magnitude, i.e. BX~Dra, V471~Tau, DP~Leo, QS~Vir and RU~Cnc. In principle, all systems show signs of strong magnetic activity.

\citet{Park2013} found strong changes in the light curves of BX~Dra, which can only be explained by large spots. Its coronal activity, however, cannot be examined due to the large distance of $230~\pc$. V471~Tau, on the other hand, exhibits photometric variability, flaring events and H$\alpha$ emission as well as a strong X-ray signal (\citet{Kaminski2007} and \citet{Pandey2008}).\\

The X-ray flux of DP~Leo has been studied extensively, e.g. by \cite{Schwope2002} and the magnetic activity of QS~Vir could be detected via Ca~II~H\&K emission, Doppler Imaging (\cite{Ribeiro2010}) and coronal emission (\cite{Matranga2012}. RU Cnc is a known ROSAT All Sky Survey (RASS) source (\cite{Zickgraf2003}).

In general, we note that it is very likely for the secondaries in these systems to show magnetic activity, as the Red Dwarfs are fully convective and rapidly rotating, due to the tidal locking to the primary star. The stars therefore likely fulfill the conditions to drive a dynamo and produce magnetic fields. The question is thus whether the activity can drive sufficiently large changes in the quadrupole moment to explain the eclipsing time variations. At least for 8 of the systems in our sample, the latter appears difficult on energetic grounds.

\subsection{Further implications}
Employing a detailed model for the Applegate mechanism in eclipsing binaries, we have checked here whether quadrupole moment variations driven by magnetic energy can explain the observed eclipsing time variations in a sample of PCEB systems. We found that at least in 8 of these systems, this possibility can be ruled out.

However, this does not mean that these systems are not magnetically active, it only implies that magnetic activity is not the only or main cause of the observed period time variations. For instance in NN~Ser, the required energy to drive the Applegate process exceeds the available energy by about a factor of $57$. Considering that the period time variations scale proportionally to the available energy, an Applegate mechanism could nevertheless contribute to additional scatter in the eclipsing time variations at a level of $\lesssim1\%$. 
While this effect may be neglected in the case of NN Ser, it may play an important role in other systems with Applegate energies closer to the energy provided by the secondary. It is therefore necessary to further investigate their possible contribution and to distinguish the latter from the potential influence of a companion in order to calculate realistic fits of planetary systems.

\section{Conclusions and discussion}
\label{sec:conclusions}

In this paper, we have systematically assessed the feasibility of the Applegate model in PCEB systems. For this purpose, we have adopted the formulation by \citet{Brinkworth2006} considering a finite shell around a central core, and including the change of the quadrupole moment both in the shell and the core. As these contributions partly balance each other, the latter is energetically more expensive than the thin-shell model by \citet{Applegate1992}, i.e. it requires more energy per orbital period to drive the eclipsing time oscillations.

We apply the Brinkworth model here in different approximations, including a constant density approximation where the required energy is independent of stellar rotation, a two-zone model assuming different densities in the shell and the core, as well as a detailed numerical model where the framework is applied to realistic stellar density profiles. We show  that the two-zone model reproduces the results of the most detailed framework with a deviation of less than $25\%$. We have explored the general dependence of the required energy. In particular, the Applegate mechanism becomes energetically more feasible for smaller binary separations, as in that case, a smaller change in the quadrupole moment is sufficient to drive the observed oscillations. In addition, the mechanism becomes more feasible with increasing mass of the secondary star, as the nuclear energy production increases with stellar mass. An ideal Applegate PCEB system consists therefore of a very tight binary ($\sim0.5~\Rsol$) with a secondary star of $\sim0.5~\Msol$.

This formalism is applied to a sample of close binaries with observed eclipsing time variations,  including the PCEB sample provided by \citet{Zorotovic2013} as well as four RS~CVn binaries.


For most systems in our sample, the energy required to drive the Applegate process is considerably larger than the energy provided by the star. 
In these cases, the observed period variations cannot be explained in the context of the Applegate model. We note that the situation is similar if we consider only the $11$ PCEBs. 
Therefore, alternative interpretations such as the planetary hypothesis need to be investigated in more detail, and we also encourage direct imaging attempts as pursued by \citet{Hardy2015} particularly in those cases where the Applegate mechanism turns out to be unfeasible.

Note that our conclusions do not imply the absence of magnetic activity for binaries where the Applegate model is not able to produce the observed period variations. Rather, it only means that the magnetic activity is not strong enough to be the dominant mechanism. However, as many of these systems have rapidly rotating secondaries with a convective envelope, we expect signs of dynamo activity, which can contribute to the period time variations on some level. 
Assuming a contribution scaling linearly with the relative Applegate threshold energy, the Applegate process might provide a significant additional scatter that needs to be taken into account when inferring potential planetary orbits from the observed data. 

\begin{acknowledgements}
MV and RB gratefully acknowledge funding from the Deutsche Forschungsgemeinschaft via Research Training Group (GrK) 1351 \textit{Extrasolar Planets and their Host Stars}. MV thanks for funding from the \textit{Studienstiftung des Deutschen Volkes} via a travel grant. We thank Klaus Beuermann, Stefan Dreizler, Rick Hessman, Ronald Mennickent and Matthias Schreiber for stimulating discussions on the topic. We also would like to thank the referee Ed Devinney for helpful comments that improved our manuscript. 
\end{acknowledgements}

\bibliography{astro-marcel.bib}
\bibliographystyle{aa}

\Online

\end{document}